\documentclass[a4paper,12pt]{article}
\usepackage[left=2.5cm,right=2.5cm,top=2.5cm,bottom=2.5cm]{geometry}
\usepackage{comment}
\usepackage{enumitem}
\usepackage{color}
\usepackage{latexsym}
\usepackage[square,numbers,sort&compress]{natbib}
\usepackage{amsmath, amsthm, amsfonts, amssymb}
\usepackage{kotex}
\usepackage{graphicx}
\usepackage{subfig}
\usepackage{mathtools}
\usepackage[symbol]{footmisc}
\usepackage{hyperref}
\usepackage{chngcntr}
\usepackage{svg}

\newtheorem{*defin}[defin]{$^{\spadesuit}$ Definition}

\newtheorem{*thm}[defin]{$^{\spadesuit}$ Theorem}

\newtheorem{*lem}[defin]{$^{\spadesuit}$ Lemma}

\newtheorem{*cor}[defin]{$^{\spadesuit}$ Corollary}

\newtheorem{*pos}[defin]{$^{\spadesuit}$ Postulate}

\newtheorem{*prop}[defin]{$^{\spadesuit}$ Proposition}

\newtheorem{*ex}[defin]{$^{\spadesuit}$ Example}

\newtheorem{*prob}[defin]{$^{\spadesuit}$ Problem}

\def\RR{\mathbb{R}} 				
\def\ZZ{\mathbb{Z}}

\def\dd{\mathrm{d}} 				

\newcommand{\abs}[1]{\left\vert#1\right\vert} 			
\newcommand{\norm}[1]{\left\Vert#1\right\Vert} 			
\newcommand{\bbr}[1]{\left[#1\right]} 					
\newcommand{\sbr}[1]{\left(#1\right)} 					
\newcommand{\bb}[1]{\mathbf#1} 							

\newcommand{\twist}[2]{\mathbf{#1}\mathbf{#2}^\top-\mathbf{#2}\mathbf{#1}^\top}

\title{A STDP-based Encoding Method for Associative and Composite Data}
\author{Hong-Gyu Yoon\footnote{hkyoon@unist.ac.kr} \;\;and\; Pilwon Kim\footnote{pwkim@unist.ac.kr, \textit{correspondence}}\smallskip  \\
        Department of Mathematical Sciences \\
        Ulsan National Institute of Science and Technology(UNIST) \\
        Ulsan Metropolitan City \\
        44919, Republic of Korea \\ }

\begin{document}

\maketitle

\begin{abstract}
Spike-timing-dependent plasticity(STDP) is a biological process of synaptic mo-\linebreak dification caused by the difference of firing order and timing between neurons. One of neurodynamical roles of STDP  is to form a macroscopic geometrical structure in the neuronal state space in response to a periodic input  \cite{susman2019stable, yoon2021stdpbased}. In this work, we propose a practical memory model based on STDP which can store and retrieve high dimensional associative data. The model combines STDP dynamics with an encoding scheme for distributed representations and is able to handle multiple composite data in a continuous manner. 
In the auto-associative memory task where a group of images are continuously streamed to the model, the images are successfully retrieved from an oscillating neural state whenever a proper cue is given. In the second task that deals with semantic memories embedded from sentences, the results show that words can recall multiple sentences simultaneously or one exclusively, depending on their grammatical relations. 

\end{abstract} \bigskip

Spike-timing-dependent plasticity(STDP) is a biological process of synaptic modification according to the order of pre- and post-synaptic spiking within a critical time window  \cite{bliss1993synaptic, bi1998synaptic, caporale2008spike}. 
In our separate work \cite{yoon2021stdpbased}, we analyzed an STDP-based neural model and showed that the model can associate multiple high-dimensional memories to a geometric structure in the neural state space which we call a \emph{memory plane}. When exposed to repeatedly occurring spatio-temporal input patterns, the neural activity based on STDP transforms the patterns into the corresponding memory plane. Further, the stored memories can be dynamically revived with macroscopic neural oscillations around the memory plane if perturbed by a similar stimulus. 

The presence and the function of the memory plane in the neural networks have caught attention in \cite{susman2019stable}, where it has been proposed that STDP can store transient inputs as imaginary-coded memories. In this work, we further emphasize a practical aspect of the memory plane, showing that it can play a central role in storing, retrieving, and manipulating structured information. Using the theoretical works in \cite{yoon2021stdpbased}, we intend to integrate an analytic and an implementation level description of the neural memory process based on the memory plane that is capable of handling high dimensional associative data. 

In this work, we propose that a STDP-based memory model, combined with a proper encoding scheme, can store and retrieve a group structured information in the neuronal state space. Among  a number of schemes for encoding compositional structure that have been proposed over the last few years, we adopt Tensor Product Representation(TPR) \cite{smolensky1990tensor}. TPR is a general method for generating vector-space embeddings of internal representations and operations, which prove to contain a variety of structural information such as lists of paired items, sequences and networks.

We show that the STDP-based memory model with TPR can naturally provide a mechanism for segmenting continuous streams of sensory input into representations of associative bindings of items: first, we demonstrate an auto-associative memory task with a group of images. While the images are sequentially streamed into the system for storage, the corresponding information is internally stored in the connectivity matrix. Then the whole gorup of the images can be dynamically retrieved from the oscillating neural state, when the system is perturbed by a memory cue which is similar to any of the original images. 
In the second task for semantic manipulation, we use multiple semantic vectors to represent a sentence as a composite of words. Once several sentences are stored in the system via such semantic vectors, a single word can recall multiple sentences simultaneously or one exclusively, depending on their grammatical relations. This implies that the proposed method provides an alternative bio-inspiring approach to process multiple groups of associative data with composite structure.

\section*{Methods}
\subsection*{STDP-based memory model}
Our work follows the framework of standard firing-rate models \cite{dayan2003theoretical, susman2019stable}. We set the differential equation for the neural state as
\begin{equation}\label{firing_eq}
\dot{\bb{x}}  = -\bb{x} + \bb{W}\phi(\bb{x}) + \bb{b(t)},
\end{equation}
where $\bb{x} = \bbr{x_1\;\cdots\;\;x_N}^\top\in\RR^N$ is the state of $N$ neuronal nodes and $\bb{W}=(W_{ij})\in\RR^{N\times N}$ is a connectivity matrix  with $W_{ij}$ corresponding to the strength of synaptic connection from node $j$ to $i$. Here  $\phi$ is a regularizing transfer function and $\bb{b}(t)$ is a memory input. 

The mechanism of STDP can be formulated as \cite{kempter1999hebbian} 
\begin{align}
\dot{W}_{ij}(t) & = -\gamma W_{ij}(t) + \rho\underbrace{\left(\int_{0}^{\infty}K(s)\phi(x_j(t-s))\phi(x_i(t))\;\dd s\right.}_{\text{pre- to post- firing}}\notag \\
& \quad\quad\quad\quad\quad\quad\quad\quad\quad\quad+ \underbrace{\left.\int_{0}^{\infty}K(-s)\phi(x_j(t))\phi(x_i(t-s))\;\dd s\right)}_{\text{post- to pre- firing}}, \label{STDP}
\end{align}
where $K$ is a temporal kernel. The parameters $\gamma$ and $\rho$ are the decaying rate of homeostatic plasticity and the learning rate, respectively. For analytic simplicity, we use $\phi(\bb{x}) = \bb{x}$ and a Dirac-delta kernel $K(s)$ defined as
\begin{equation}\label{kernel}
K(s) :=\begin{dcases}
\delta(s-s_0) & s > 0 \\
-\delta(s+s_0) & s \le 0,
\end{dcases}
\end{equation}
with $s_0>0$. After simplifications, the main model becomes

\begin{equation}\label{system}
\begin{dcases}
\dot{\bb{x}} = -\bb{x} + \bb{W}\bb{x} + \bb{b(t)} \\
\dot{\bb{W}} = -\gamma\bb{W}+\rho\sbr{ \twist{\bb{x}}{\bb{x}_\tau}},
\end{dcases}
\end{equation}
where $\bb{x}_\tau = \bb{x}(t-\tau)$ stands for the delayed synaptic response. 



We use the memory input in the form of a sequential harmonic pulse as
\begin{equation}\label{memory_input1}
\bb{b}(t) = \sum_{i=1}^{n}\sin(\omega t-\xi_i)\bb{m}_i,
\end{equation}
where $\bb{m}_1,\dots,\bb{m}_n\in\RR^N$ are memory representations to be stored.  
Here $\omega$ stands for the frequency of neural oscillations and $\xi_i$, $i=1,\dots,n$ stands for the sampling time for each component. The trajectory of the memory input $\bb{b}(t)$ in (\ref{memory_input1}) is periodic and embedded in a 2-dimensional plane $S \subset\RR^N$ which we call a memory plane with respect to the memory representations $\bb{m}_1,\dots,\bb{m}_n$. While the memory representations are distributed in the high dimensional neural state space $\RR^N$, the memory plane $S$ tends to be located in close proximity to the memory representations under a suitable condition \cite{yoon2021stdpbased}. Further, the system (\ref{system}) has a asymptotically stable solution $(\bb{x}^*, \bb{W}^*)$ that consists of a periodic solution $\bb{x}^*(t)$
 on the memory plane $S$ and a constant connectivity matrix $\bb{W}(t)=\bb{W}^*=\alpha(\twist{\bb{v}}{\bb{u}})$ for some vectors $\bb{u}$ and $\bb{v}$ in $S$.
 We will see below that the matrix $\bb{W}^*$ contains the essential information to retrieve the memory representations $\bb{m}_1,\cdots,\bb{m}_n$. This implies, a convergence of $\bb{x}(t)$ to a certain periodic oscillation, $\bb{x}^*(t)$, is a sign that the memories are stored in $\bb{W}$ in a distributed way. 

\begin{figure}[ht!]
\centering
  \includegraphics[width=1\textwidth]{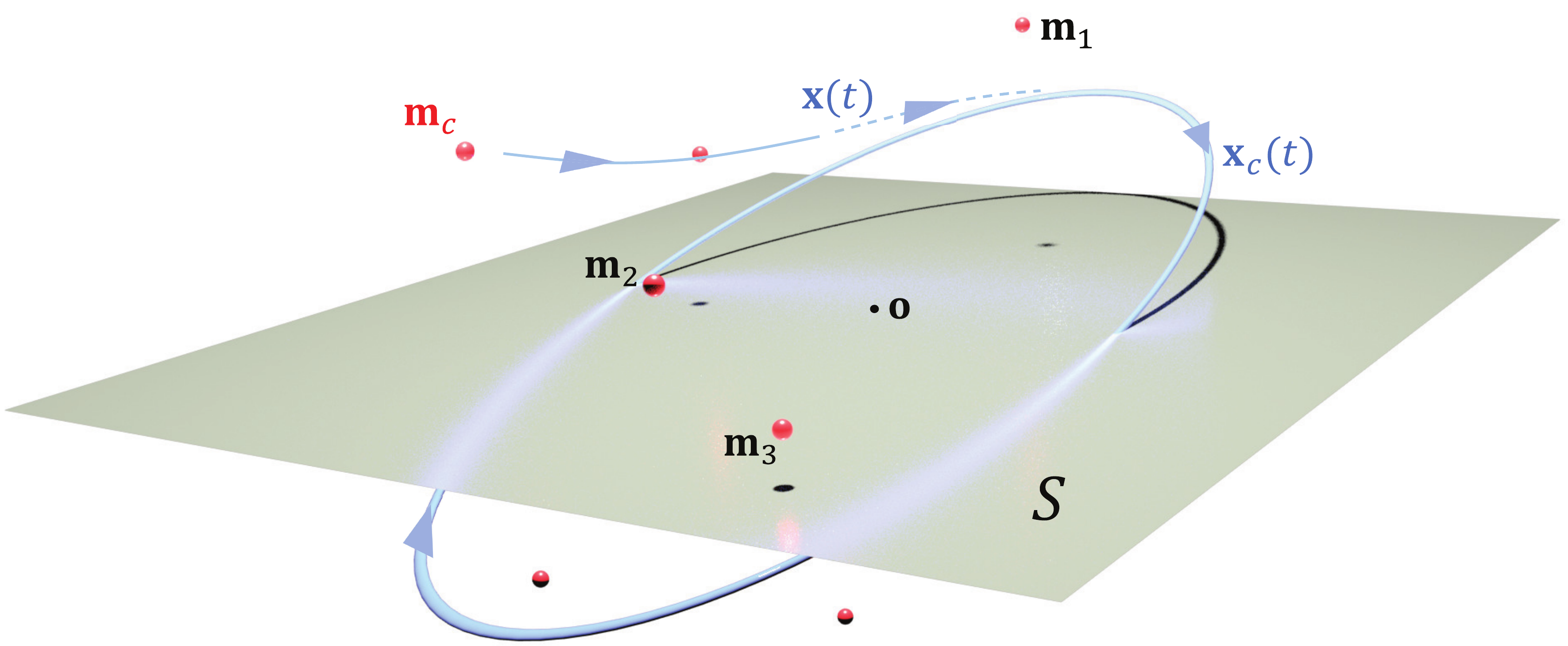}
  \caption{\small Graphical illustration of the memory representations $\bb{m}_1, \dots, \bb{m}_n$ and the corresponding memory plane $S$. The memory plane $S$ is located in the subspace spanned by the memory representations  and is shown to be close enough to them. A periodic orbit close to $S$ can be used for efficient memory retrieval.} \label{memory_plane}
\end{figure}

In the retrieval phase, we set $\gamma=\rho=0$ in Eq. (\ref{system}) and use the connectivity matrix $\bb{W}=\bb{W}^*$ as 
\begin{equation}\label{retrieval_eq}
\dot{\bb{x}}= -\bb{x}+\bb{W}^*\bb{x}+\bb{b}(t).
\end{equation}
where
\begin{equation}\label{memory_input2}
\bb{b}(t) = \sin\omega t\,\bb{m}_c,\quad\bb{m}_c\in\RR^N.
\end{equation}
Here $\bb{m}_c$ is the representation of memory cue.  
The retrieval system (\ref{retrieval_eq}) has also a asymptotically stable periodic solution, say $\bb{x}_c(t)$. One can show that the trajectory of the periodic solution $\bb{x}_c(t)$ is closely located to $S$ if the memory cue $\bb{m}_c$ is relevant to any of the memory representations $\bb{m}_1,\cdots,\bb{m}_n.$
Hence, as a neural state $\bb{x}(t)$ is atrracted to $\bb{x}_c(t)$,  it is also bound to circle around the memory representations $\bb{m}_1,\dots,\bb{m}_n$.
%
%
Fig. \ref{memory_plane} shows that  the neural oscillation occurs near the memory plane and therefore in proximity to all the memory representations, when $\bb{m}_c$ is given close to one of them.

\subsection*{Encoding/Decoding associative data with tag vectors}
Before storing and retrieving  specific associative data in (\ref{system}) and (\ref{retrieval_eq}), respectively, we first need to properly encode them into memory representations $\bb{m}_1,\dots,\bb{m}_n$.
It is reasonable to assume that the cognitive systems do not simply receive external inputs in a passive way, but rather actively pose them in the neural state space on acceptance. Suppose $\bb{f}_1, \dots,  \bb{f}_n\in\RR^{D}$  are a series of the external inputs from the environment containing high dimensional associative information. We use a set of internal tags $\bb{r}_1,  \dots, \bb{r}_m\in\RR^{K}$ to mark what classes the corresponding external inputs belong to. They may be used to indicate the order of sequence for the events (the first, the second, \ldots, the last) if the input is streamed through sequential observations, or types of sensors (visual, auditory, olfactory, tactile) if the input is a combination of senses, or the sentence elements (subject, predicate, object, modifier) if the input is a sentence composed of words. Such internal tags $\bb{r}_1,  \dots, \bb{r}_m$ can be formulated as low dimensional orthonormal vectors. 

Following the vector embedding method for structured information \cite{smolensky1990tensor}, we use the tensor product to encode a raw data $\bb{f}_i$ into a memory representation $\bb{m}_i$ as 
\begin{equation}\label{tensor}
\bb{m}_i=\bb{f}_i\otimes\bb{r},
\end{equation}
where $\bb{r}$ is the tag vector corresponding to the raw data $\bb{f}_i$.  
If $\bb{r}$ is of unit length, the original data can be exactly decoded from the memory representations by applying the right dot product of tagging vector as
\begin{equation}\label{eq_recovery}
\bb{m}_i\cdot\bb{r} = (\bb{f}_i\otimes\bb{r})\cdot\bb{r}=\bb{f}_i(\bb{r}\cdot\bb{r})=\bb{f}_i.
\end{equation}
We say a neural state $\bb{x}\in\RR^{N}$ to be \emph{retrievable} if $\bb{x}$ is a linear combination of the memory representations $\bb{m}_1,\cdots,\bb{m}_n$. For a retirevable state $\bb{x}(t)=\sum_{j=1}^{n}c_j\bb{m}_j$,  
a selective recovery of the original $\bb{f}_i$ is possible by applying the right dot product with the corresponding tag vector as
\begin{equation}\label{eq_tensor_reconst}
\begin{split}
\bb{x}\cdot\bb{r}_i=\sbr{\sum_{j=1}^{n}c_j\bb{m}_j}\cdot\bb{r}_i & =  c_i\bb{f}_i.
\end{split}
\end{equation}
It can be shown that all the points on the memory plane $S$ with respect to $\bb{m}_1,\dots,\bb{m}_n$ are retrievable.

Let us describe how the encoding/decoding scheme is combined with the above memory models. In the storage phase, we encode the data $\bb{f}_1, \dots,  \bb{f}_n$ into the memory representations $\bb{m}_1, \dots,  \bb{m}_n$ using the tag vectors in (\ref{tensor}), and run the system (\ref{system}) with the memory input $\bb{b}(t)$ in (\ref{memory_input1}). To retrieve the original data, we run the retrieval system (\ref{retrieval_eq}) with a certain cue $\bb{m}_c$. 
We wait until $\bb{x}(t)$ converges to an oscillating trajectory, $\bb{x}_c(t)$, then try to evaluate $\bb{x}(t)\cdot \bb{r}$ for retrieval.
It has been shown \cite{yoon2021stdpbased} that $\bb{x}_c(t)$ always has intersections with $S$ which are therefore retrievable. However, since the neural state space $\RR^N$ is high dimensional, an irrelevant choice for $\bb{m}_c$ leads to a trivial intersection near the origin which yields no meaningful retrieval.
To the contrary, if $\bb{m}_c$ is close to any of $\bb{m}_1, \dots,  \bb{m}_n$, then $\bb{x}(t)$ converges to a periodic solution $\bb{x}_c(t)$ which is almost embedded $S$ as illustrated in Figure1.
Indeed, one can show that if $\bb{m}_c$ is a scalar multiple of one of the memory representations, the corresponding $\bb{x}_c(t)$ is completely embedded in $S$ and is therefore retrievable for all $t$.

In the following section, we show through numerical tests for storage and retrieval that the STDP-based model (\ref{system}) with the tensor product encoding can naturally provide a neural mechanism for segmenting continuous streams of sensory input into representations associative bindings of items.

\begin{figure}[ht!]
\centering
  \includegraphics[width=0.85\textwidth]{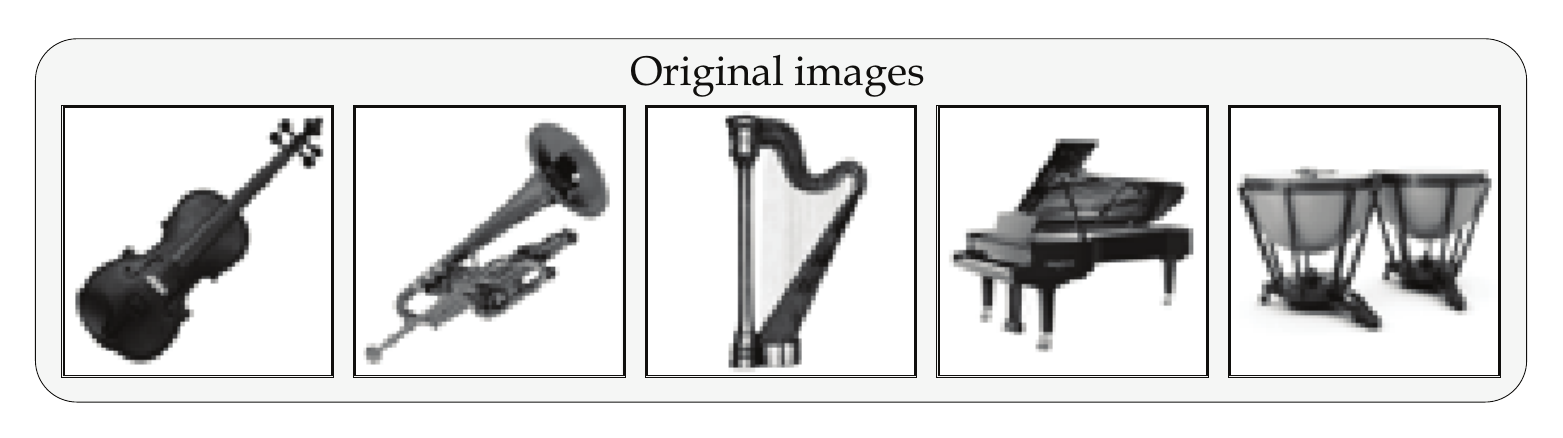}
  \caption{\small  Grayscale images of $64\times 64$ pixels displaying classical orchestral instruments that are used for the memory input vectors $\bb{f}_1,\dots,\bb{f}_5$. } \label{image_group}
\end{figure}

\subsection*{Results}

\subsubsection*{Retrieval of grouped images}

We first demonstrate an auto-associative memory task that involves a group of images. This task uses five $64\times 64$  grayscale images of classical orchestral instruments in Fig. \ref{image_group}. The images are translated into external input vectors $\bb{f}_i$, $i=1,\dots,5$ in $\RR^{64^2}$ and are combined into the memory representations  as $\bb{m}_i=\bb{f}_i\otimes\bb{r}_i$, $i=1,\dots,5$. Here the tag vectors $\bb{r}_i$, $i=1,\dots,5$ are orthonormal in $\RR^5$ and used as a placeholder for each image.


\begin{figure}[ht!]
\centering
  \includegraphics[width=1\textwidth]{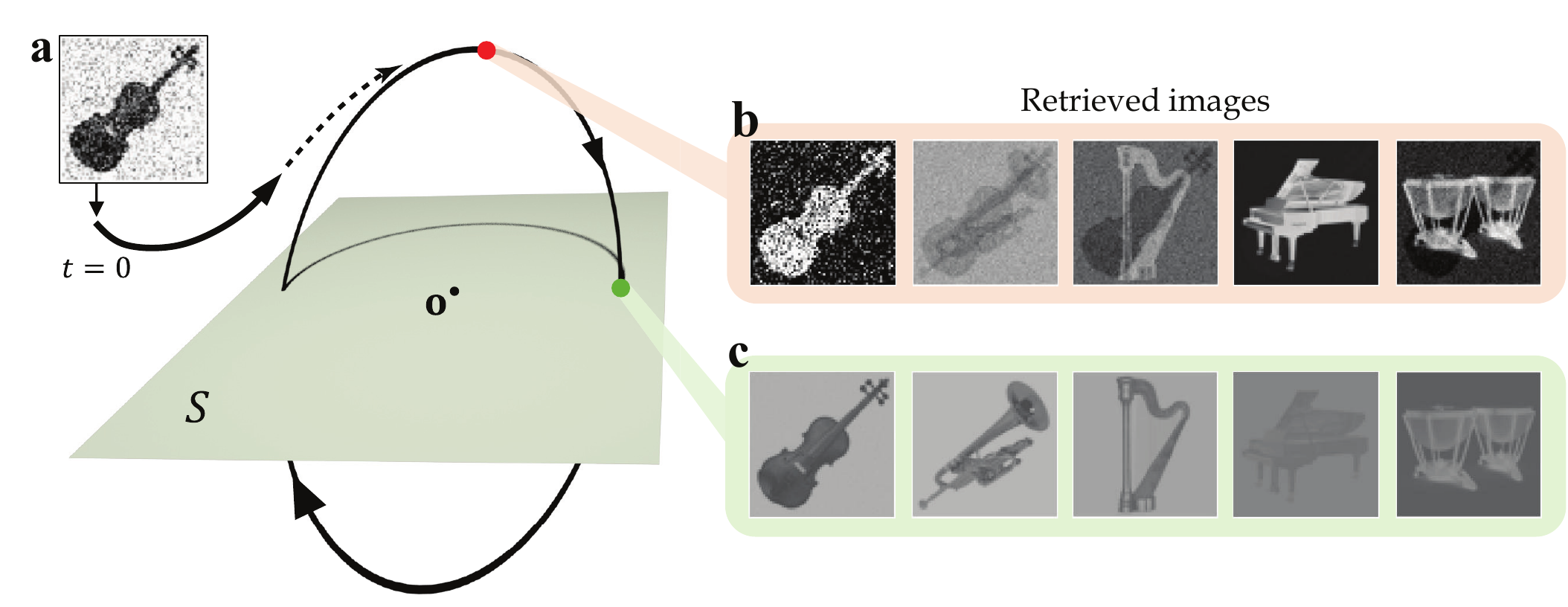}
  \caption{\small Auto-associative memory retrieval from a contaminated cue.
\textbf{(a)} The nosisy cue $\bb{m}_c=\widetilde{\bb{f}_1}\otimes\widetilde{\bb{r}_1}$ is generated from $\widetilde{\bb{f}_1}=\sqrt{1-\alpha^2}\bb{f}_1+\alpha\boldsymbol\zeta$ and $\widetilde{\bb{r}_1} = \sqrt{1-\beta^2}\bb{r}_1+\beta\boldsymbol\eta,$ where $\boldsymbol\zeta$, $\boldsymbol\eta$ are Gaussian noise following  $\mathcal{N}_{\RR^{64^2}}(0,\norm{\bb{f}_1})$ and $\mathcal{N}_{\RR^5}(0,1),$ respectively. The parameters are $\alpha = 0.25$ and $\beta = 0.2$. \textbf{(b)} Snapshot of the retrieved images at the farthest point (red dot) from memory plane $S$. \textbf{(c)} Snapshot of the retrieved images at the intersection (green dot) of the orbit and the memory plane $S$. The timing of intersection $t=t^\dagger>0$ can be analytically determined as $t^\dagger = (\tan^{-1}\omega + n\pi)/\omega$, $n\in\ZZ$ \cite{yoon2021stdpbased}.  } \label{retrieval_illust}
\end{figure}

\begin{figure}[!ht]
\centering
  \includegraphics[width=1\textwidth]{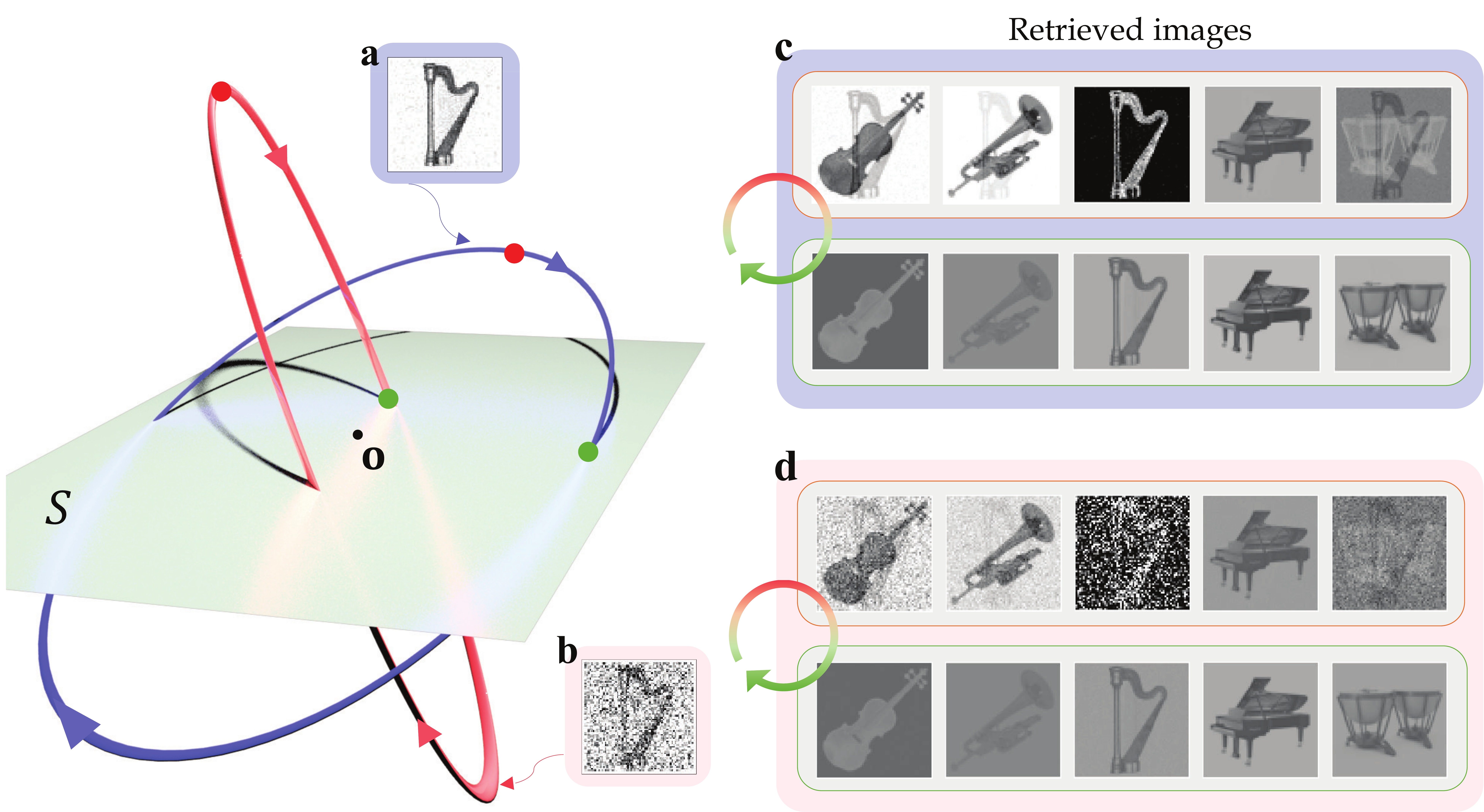}
  \caption{\small Comparison of retrieval quality according to the noise level in the cue. 
  \textbf{(a)} The less noisy cue. The cue is generated in the same way as in Fig. \ref{retrieval_illust} except using $\bb{f}_3$ and $\bb{r}_3$ instead of $\bb{f}_1$ and $\bb{r}_1$. Gaussian noise are $\mathcal{N}_{\RR^{64^2}}(0,\norm{\bb{f}_3})$ and $\mathcal{N}_{\RR^5}(0,1)$, respectively. The parameters are $\alpha = 0.1$ and $\beta=0.2$. \textbf{(b)} The severely contaminated cue with $\alpha = 0.7$. \textbf{(c)} Snapshots of the retrieved images from the less noisy cue in (a), taken at the farthest point from $S$ (top row) and at the intersection (bottom row). \textbf{(d)} Snapshots of the retrieved images from the more noisy cue in (b), taken at the farthest point from $S$ (top row) and at the intersection (bottom row).} \label{noise_level}
\end{figure}

\begin{figure}[!ht]
\centering
  \includegraphics[width=0.87\textwidth]{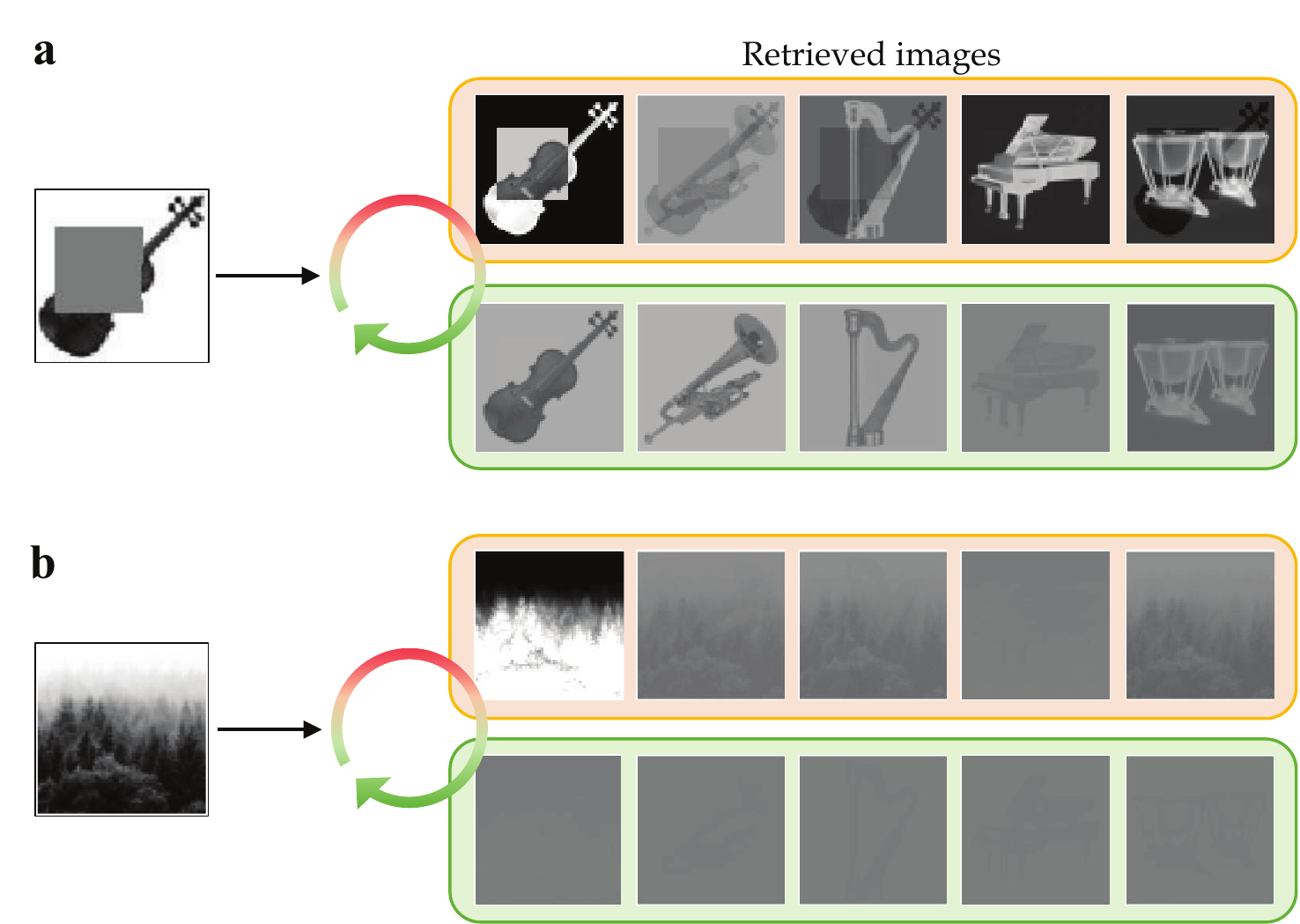}
  \caption{\small \textbf{(a)} Result of retrieval from a partially obstructed cue. Snapshots of the retrieved images are taken at the farthest point from $S$ (top row) and at the intersection of the orbit and the memory plane (bottom row). \textbf{(b)} Result of retrieval from an irrelevant cue. Snapshots of the retrieved images are taken at the farthest point from $S$ (top row) and at the intersection of the orbit and the memory plane (bottom row). In both (a) and (b), we used the noisy tag vector $\widetilde{\bb{r}_1}$ used in Fig. \ref{retrieval_illust} for retrieval. } \label{blocked_irrel}
\end{figure}

For numerical simulations in this article, the modified Euler's method for delay equations has been universally used. In the first memory task, each $64\times 64$ image is translated to a vector $\bb{f}_i$ as follows: every pixel is mapped to a value in $[-\sigma,\sigma], \sigma>0$, depending on its brightness (pure black to $-\sigma$ and pure white to $\sigma$ linearly). Then the resulted $64\times 64$ matrix is flattened to a vector $\bb{f}_i$. In the storage phase, $\sigma=0.02$ was used to maintain the magnitude for $\bb{f}_i$ and $\bb{m}_i$ at an appropriate level. Reconstructing the image from the vector can be done by performing the procedure in reverse order.
The storage phase was proceeded for 40 seconds with the integration step size $\Delta t=0.1$ and $\xi_i=\frac{\pi}{5}(i-1)$ (5-evenly sequenced points on $[0,\pi]$). The used parameters are $\omega = 1.5$, $\gamma = \rho = 0.5$, and $\tau = \frac{\pi}{3}$, respectively. Stable convergence of connectivity to $\bb{W}^*$ is well achieved, when the initial condition $\bb{x}(0)$ and $\bb{W}(0)$ are appropriately small.  
The retrieval phase was proceeded for 15 seconds with $\Delta t=0.01$ for appropriately small $\bb{x}(0)$. In Fig. \ref{retrieval_illust}, \ref{noise_level}, and \ref{blocked_irrel}, the brightness threshold $\sigma$ was adjusted to $0.005$ for clear visibility, since the magnitude of the retrieved images are relatively small compared to original ones. Thus, any element of $\bb{x}\cdot\bb{r}_i$ having value outside $[-0.005\;\;0.005]$ is developed to a pixel of just pure black or white.

Fig. \ref{retrieval_illust} depicts the numerical simulation for the retrieval phase.  For a better understanding of the process, a graphic illustration of the memory plane and the initial memory cue is given with the actual data. When the neural state $\bb{x}(t)$ in Eq. (5) is continually perturbed by a noisy copy of one of the original images (violin), it approaches the memory plane $S$. Once the $\bb{x}(t)$ converges to a limit cycle around $S$  as stated in Theorem 2,  the external input $\bb{f}_1,\dots,\bb{f}_5$ can be reproduced by applying the tag vectors to $\bb{x}(t)$. In Fig. \ref{retrieval_illust}, we display two snapshots of the retrieved images obtained at two points on the orbit: Fig. \ref{retrieval_illust}b is taken at the farthest from $S$ and Fig. \ref{retrieval_illust}c is at the intersection. It is notable that the retrieved images continuously oscillate, developing week/strong and positive/negative images in turns. Such flashing patterns are generally different from image to image and are affected by the sequential order of the memory representations in Eq. (\ref{memory_input1}) in the storage phase. Furthermore, due to the orthogonality of the tag vectors, the perfect images are acquired on the time instance when $\bb{x}(t)$ penetrates $S$.


Fig. \ref{noise_level} shows that the quality of the retrieved images depends on how close the memory cue is to the original image input. The cue with low-level noise in Fig. \ref{noise_level}a leads to the orbit (blue) close to the memory plane $S$, producing the images of decent quality in Fig. \ref{noise_level}c. However, if the cue is more contaminated with noise as in Fig. \ref{noise_level}b,  $\bb{x}(t)$ approaches $S$ at a relatively larger angle, making a stretched narrower elliptical orbit (red) that periodically gets far from $S$. Although the orbit from the severely contaminated cue still passes through the memory plane, it only does near the origin, providing relatively feeble images during a short time.

The retrieval can be performed with an incomplete cue.  
In Fig. \ref{blocked_irrel}a, the images are recalled from the partially obstructed cue. The original images can be recovered at a decent level, especially when $\bb{x}(t)$ passes through the memory plane $S$.
Fig. \ref{blocked_irrel}b displays that an irrelevant cue (forest) fails to retrieve the original memory inputs. Indeed, it can be shown that a completely irrelevant cue results in a one-dimensional periodic orbit that keeps penetrating the memory plane back and forth just at the origin. 

\subsubsection*{Multiple groups of memory with composite structure}

This section deals with applications of the model to more complex associative memory.
Suppose we have multiple groups of memory representations and have stored each group in the form of the memory plane using the system in Eq. (\ref{system}). We are especially interested in the case where some memory representations belong to multiple groups. The following questions naturally arise: 1) Can the common memory component retrieve the corresponding multiple groups together? 2) Can a single memory group be selected by adding a further memory component in the cue? These questions are potentially related to high-level inference on memory.

We also focus on compositional structure of memory representations created by the tag vectors.  Memory inputs in this section are words and are collectively provided in the form of a sentence. 
We assume that each tag vector stands for the sentence element (subject, predicate, object, modifier) and is naturally bound to a word according to the role of the corresponding word in the sentence. Being activated by such a sequential stream of words, the system in Eq. (\ref{system})  forms the memory plane which can be referred to as the encoding of the sentence.

\begin{figure}[ht!]
\centering
  \includegraphics[width=0.85\textwidth]{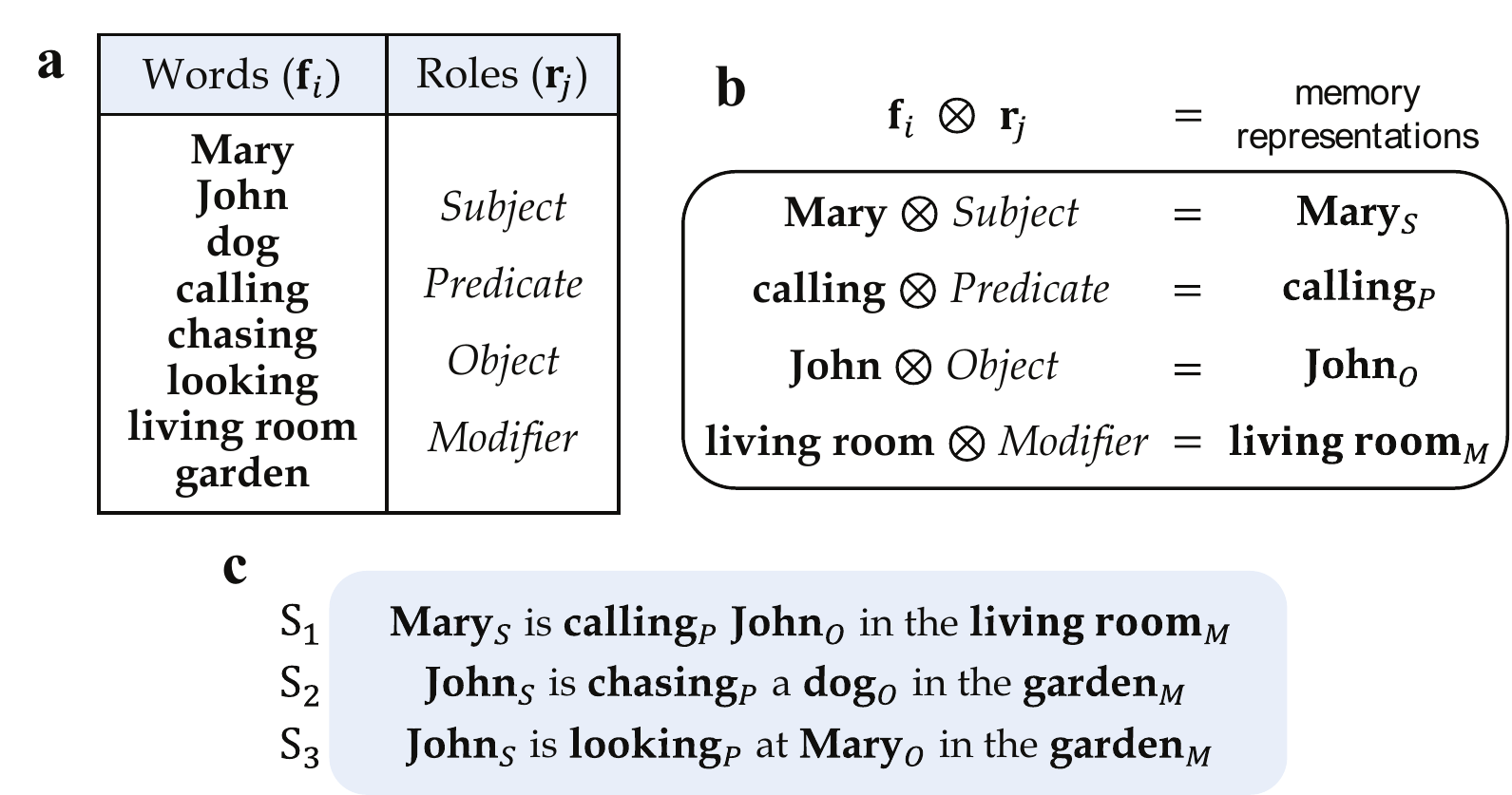}
  \caption{\small \textbf{(a)} List of words and roles used for the external input $\bb{f}_i$ and the tag $\bb{r}_j$. \textbf{(b)} Descriptive explanation on constructing memory representations. \textbf{(c)} Three sentences $\text{S}_1$, $\text{S}_2$ and $\text{S}_3$ generated by grouping memory representations. Note that the words \textbf{John}, \textbf{Mary} and \textbf{garden}  appear several times in different contexts.}\label{composite_input}
\end{figure}

For the simulation of semantic memory, we use three sentences composed of 8 words. Every word appearing in the sentence has one of the 4 roles (sentence elements). The vocabulary of the words and the roles are listed in Fig. \ref{composite_input}a. We simply use arbitrarily chosen orthonormal sets $\{\bb{f}_i\}_{i=1}^8$ for the words and $\{\bb{r_j}\}_{j=1}^4$ for the roles, respectively.  
Fig. \ref{composite_input}b shows a couple of examples for memory representations each of which is a binding of a word and a role.  Here the subindex on the right-hand side is used to express the corresponding role for the word. 
Our goal is to store the semantic information of sentences through Eq. (\ref{system}) with the memory input $\bb{b}(t)$ in Eq. (\ref{memory_input1}). There are three sentences $\text{S}_1, \text{S}_2$ and $\text{S}_3$ listed in Fig. \ref{composite_input}c that we use as the memory input in the simulation. 
Note that word \textbf{John} appears three times in the sentences, once in $\text{S}_1$ as an object, and twice in $\text{S}_2$ and $\text{S}_3$ as a subject. Similarly, the words  \textbf{Mary} and \textbf{garden} occur twice in a different context.

The memory connectivity $\bb{W}_k^*$, $k=1,2,3$ are obtained from separate single group learning on the sentences $\text{S}_k$, $k=1,2,3$, respectively. We then set the combined memory connectivity for three sentences, i.e., $\bb{W}^*=\bb{W}^*_1+\bb{W}^*_2+\bb{W}^*_3$ for the collective retrieval phase.  We adopt the function  
\begin{equation}\label{measure}
P^i_j(t):=\int_{t_0}^{t}\abs{\bb{f}_i^\top(\bb{x}(t)\cdot\bb{r}_j)}\dd s,\quad i=1,\dots,8,\;j=1,\dots,4,
\end{equation}
to measure how close the retrieved quantity is to the word $\bb{f}_i$ as the role $\bb{r}_j$.

In the following sematics tasks, the retrieval was proceeded for 30 seconds with $\Delta t = 0.01$ for appropriately small $\bb{x}(0)$. Multiple cues such as $\textbf{John}_S+\textbf{Mary}_O$ in Fig. \ref{1to2_2to1}d are implemented by assigning each cue to its original sampling time through a harmonic pulse. In other words, the combined cue $\textbf{John}_S+\textbf{Mary}_O$ is implemented as $\bb{b}_c(t) = \sin(\omega t-\xi_1)(\bb{f}_2\otimes\bb{r}_1)+\sin(\omega t-\xi_3)(\bb{f}_1\otimes\bb{r}_3)$.

\begin{figure}[ht!]
\centering
  \includegraphics[width=1\textwidth]{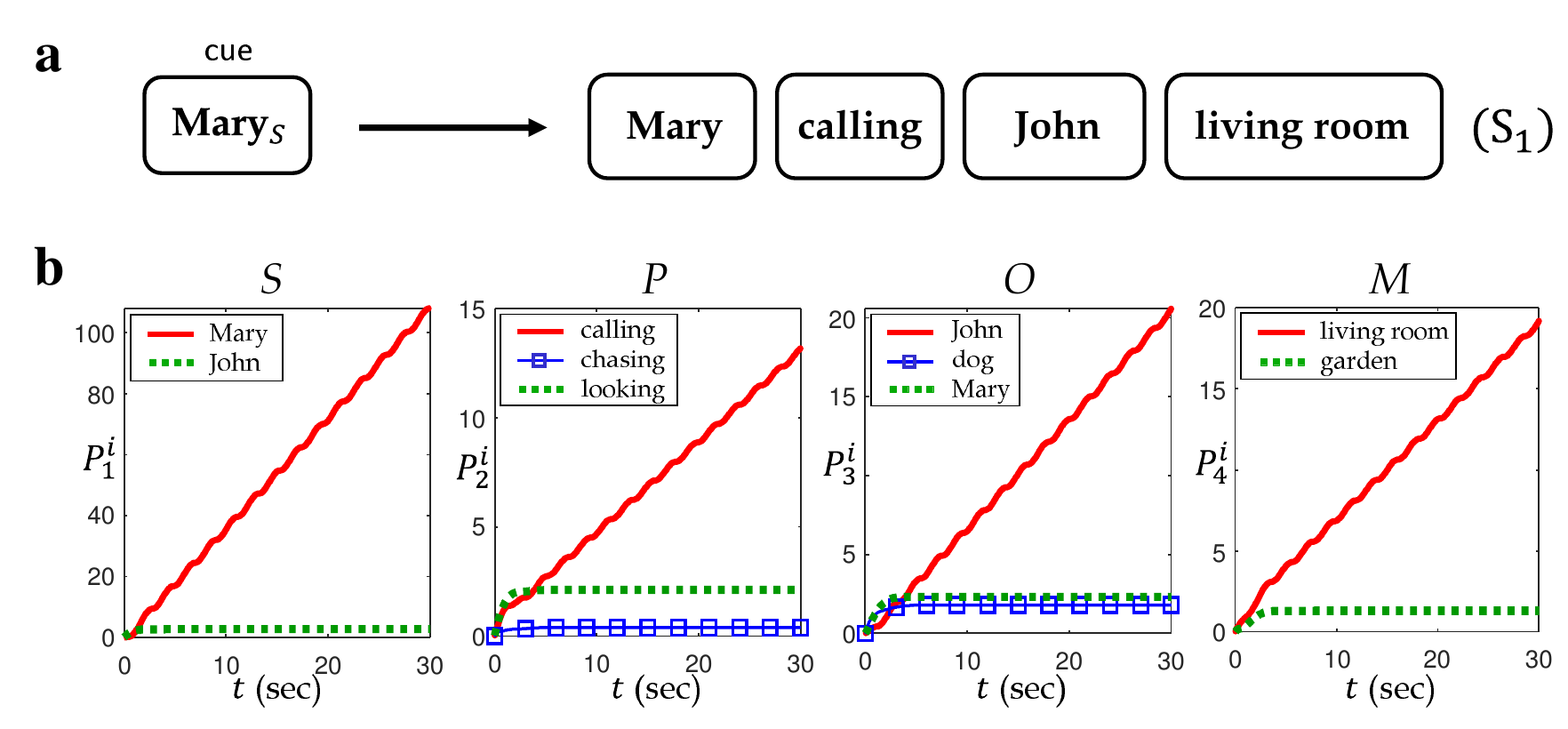}
  \caption{\small  \textbf{(a)} Expected result of retrieval when the cue $\textbf{Mary}_S$ given.  \textbf{(b)} The numerical result of retrieval by the cue $\textbf{Mary}_S$. Dominant increasing values of $P_i^j(t)$ are colored red, which turned out to correspond to $\text{S}_1$. } \label{among_three}
\end{figure}

\begin{figure}[ht!]
\centering
  \includegraphics[width=1\textwidth]{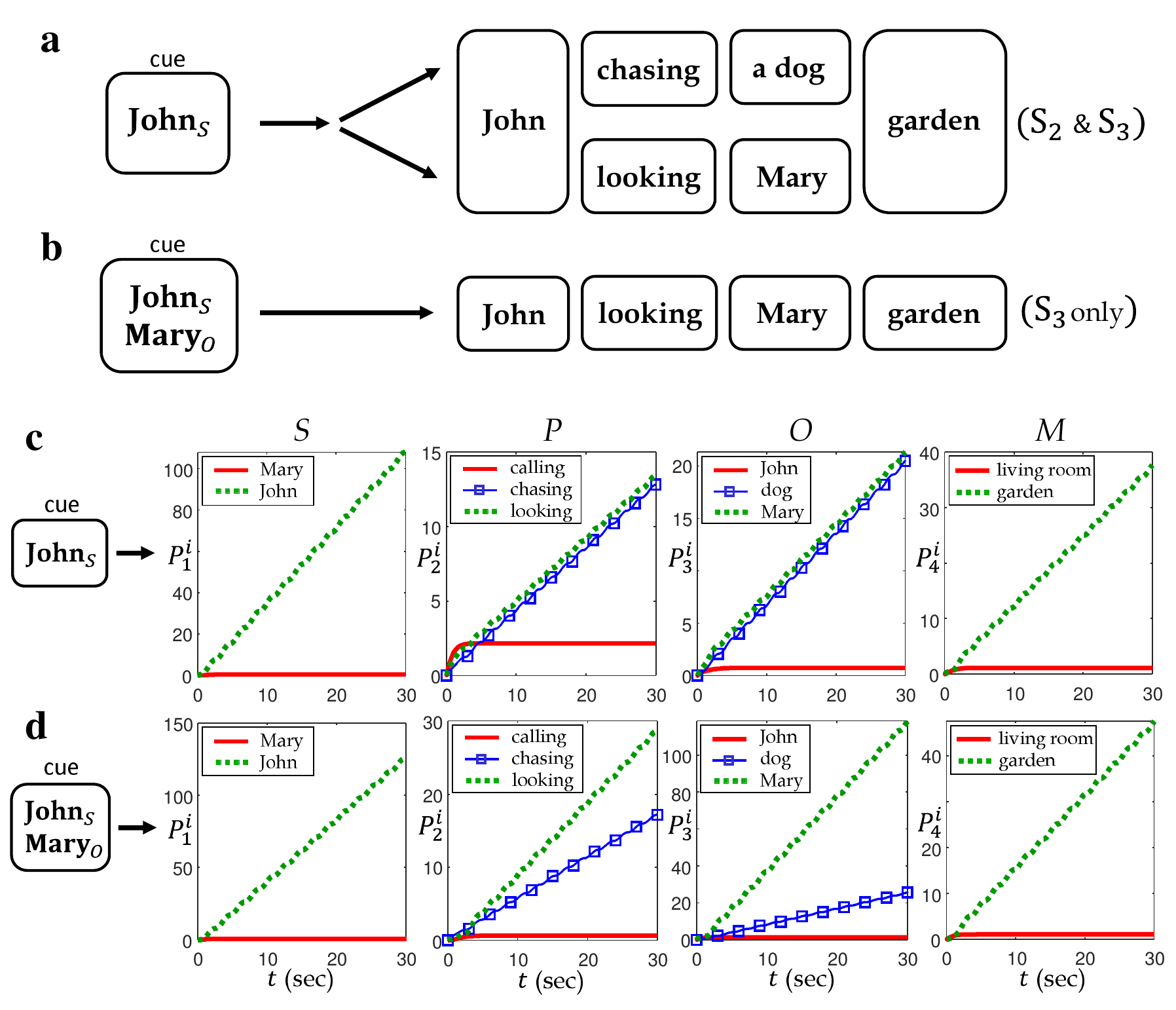}
  \caption{\small \textbf{(a)} Expected result of retrieval by a single component cue $\textbf{John}_S$. \textbf{(b)} Expected result of retrieval by multiple component cues $\textbf{John}_S$ and $\textbf{Mary}_O$. \textbf{(c)} Numerical result of retrieval by a single component cue $\textbf{John}_S$. \textbf{(d)} Numerical result of retrieval by multiple component cues $\textbf{John}_S$ and $\textbf{Mary}_O$.} \label{1to2_2to1}
\end{figure}

In the first task of multiple composite memories, $\textbf{Mary}_S$ is given as the cue. Since \textbf{Mary} occurs as a subject only in $\text{S}_1$, one can expect the retrieved result to be  $\text{S}_1$ as in Fig. \ref{among_three}a. The numerical simulation of the retrieval process turned out to agree well with this expectation. Fig. \ref{among_three}b compares the fitness of the words. 
The values of $P^i_j(t)$ in Eq. (\ref{measure}) are evaluated while $\bb{x}(t)$ is oscillating along a convergent orbit of Eq. (\ref{retrieval_eq}). If $P^i_j(t)$ keeps increasing with a large slope, the corresponding memory component $\bb{f}_i\otimes\bb{r}_j$ can be identified as a dominantly retrieved one. 
The graphs in Fig. \ref{among_three}b show that such representations are  $\textbf{Mary}_S$, $\textbf{calling}_P$, $\textbf{John}_O$ and $\textbf{living room}_M$, which are well matched to $\text{S}_1$.


The second task deals with the case with an ambiguous memory cue.
Suppose that the memory component $\textbf{John}_S$ is given as the cue.
Since it occurs in the both sentence $\text{S}_2$ and $\text{S}_3$,
it is reasonable that the retrieval result should involve all the memory representations in both sentences as in Fig. \ref{1to2_2to1}a. This may be understood in that one of the fundamental capabilities of the brain is to examine all possible memories that contain the common cue, especially when the given cue is insufficient. However, this ambiguity can be eliminated by adding further cues. For example, if $\textbf{Mary}_O$ is added as in Fig. \ref{1to2_2to1}b, the retrieval result should be narrowed down to $\text{S}_3$ due to the extra constraint. 

It turned out that the numerical simulations successfully capture the expected features of the memory retrieval process mentioned above. Fig. \ref{1to2_2to1}c shows the results from the memory cue $\textbf{John}_S$. It is notable that $\textbf{chasing}_P$ and $\textbf{looking}_P$ in the second graph simultaneously increase with the almost same slope, indicating that they are equally dominant memory representations in retrieval. This is even clearer when compared to another memory component $\textbf{calling}_P$ which is steady and negligible. The same pattern appears with $\textbf{John}_O$ and $\textbf{dog}_O$ in the third graph, both of which are dominant retrieval representations. The numerical results in Fig. \ref{1to2_2to1}d also reflect the retrieval tendency with additional memory cue. We provide the system in Eq. (\ref{retrieval_eq}) with the extended memory input in Eq. (\ref{memory_input2}) that consists of two memory representations $\textbf{John}_S$ and $\textbf{Mary}_O$. Since the newly added cue  $\textbf{Mary}_O$ confines the retrieval result to the sentence $\text{S}_3$ as in Fig. \ref{1to2_2to1}b, the memory representations in $\text{S}_2$, $\textbf{chasing}_P$ and $\textbf{dog}_O$, should be suppressed in retrieval. The second and third graphs in Fig. \ref{1to2_2to1}d show that, while $\textbf{chasing}_P$ and $\textbf{dog}_O$ increase (due to the common cue $\textbf{John}_S$), the slope is smaller than that of $\textbf{looking}_P$ and $\textbf{Mary}_O$ in $\text{S}_3$, respectively. This implies that the dominantly retrieved representations are $\textbf{John}_S$, $\textbf{looking}_P$, $\textbf{Mary}_O$ and $\textbf{garden}_M$ which are matched to $\text{S}_3$.

\section*{Discussion}

There is now substantial evidence accumulated that neural oscillations are related to memory encoding, attention, and integration of visual patterns \cite{singer1995visual, gupta2016oscillatory, rutishauser2010human}. In \cite{susman2019stable}, the idea has been proposed that memories constitute stable dynamical trajectories on a two-dimensional plane in which an incoming stimulus is encoded as a pair of imaginary eigenvalues in the connectivity matrix. We extended such an idea further through a specific memory system that can process a group of high dimensional associative data sets, by using the exact analytical relation between the inputs and the corresponding synaptic changes shown in \cite{yoon2021stdpbased}. 
Different from the Hopfield network that retrieves static single data as a fixed point, the proposed model produces neural oscillations in response to an external cue, exploring various aspects of stored multiple data sets around the memory plane.

We encode the input data with tag vectors based on the tensor representation, which has been proposed as a robust and flexible distributed memory representation \cite{hintzman1984minerva, kanerva1988sparse, humphreys1989different,smolensky1990tensor, pollack1990recursive}. This preprocess enables us to efficiently retrieve the stored data and, in addition, to deal with the composite structure in the data set. 
The ability to process associate multiple data sets with composite structures is essential in natural language understanding and reasoning. It has been shown that the proposed model can handle multiple sentences that describe distinct situations and can selectively allow the recall cue to arouse a group of associative memories according to its semantic relevance. 

From a practical perspective, our results suggest an alternative approach for a memory device. The conventional von Neumann architecture is non-scalable and its performance is limited by the so-called von Neumann bottleneck between nonvolatile memories and microprocessors. On the other hand, operating data with artificial synapses is benefiting from a parallel information process consuming a small amount of energy per synapse. Moreover, conventional digital memory systems convert the inputs to a binary code and save it in a separate storage device, likely destroying the correlation information by such physical isolation. The proposed model is based on continuous dynamical systems and provides a simple and robust approach to deal with a sequence of associative high-dimensional data. Processing data in the continuous and distributed system results in the plastic storage of the correlated information in the synaptic connections. 

\subsection*{Acknowledgements}
P. Kim was supported by National Research Foundation of Korea (2017R1D1A1B040\linebreak 32921) and H. Yoon was supported by Ulsan National Institute of Science and Technology 12(1.200052.01).

\bibliography{memory_bib.bib}

\begin{thebibliography}{10}

\bibitem{susman2019stable}
L.~Susman, N.~Brenner, and O.~Barak, ``Stable memory with unstable synapses,''
  {\em Nature communications}, vol.~10, no.~1, pp.~1--9, 2019.

\bibitem{yoon2021stdpbased}
H.-G. Yoon and P.~Kim, ``Stdp-based associative memory formation and
  retrieval,'' {\em arXiv preprint arXiv:2107.02429v2}, 2021.

\bibitem{bliss1993synaptic}
T.~V. Bliss and G.~L. Collingridge, ``A synaptic model of memory: long-term
  potentiation in the hippocampus,'' {\em Nature}, vol.~361, no.~6407,
  pp.~31--39, 1993.

\bibitem{bi1998synaptic}
G.-q. Bi and M.-m. Poo, ``Synaptic modifications in cultured hippocampal
  neurons: dependence on spike timing, synaptic strength, and postsynaptic cell
  type,'' {\em Journal of neuroscience}, vol.~18, no.~24, pp.~10464--10472,
  1998.

\bibitem{caporale2008spike}
N.~Caporale and Y.~Dan, ``Spike timing--dependent plasticity: a hebbian
  learning rule,'' {\em Annu. Rev. Neurosci.}, vol.~31, pp.~25--46, 2008.

\bibitem{smolensky1990tensor}
P.~Smolensky, ``Tensor product variable binding and the representation of
  symbolic structures in connectionist systems,'' {\em Artificial
  intelligence}, vol.~46, no.~1-2, pp.~159--216, 1990.

\bibitem{dayan2003theoretical}
P.~Dayan, L.~F. Abbott, {\em et~al.}, ``Theoretical neuroscience: computational
  and mathematical modeling of neural systems,'' {\em Journal of Cognitive
  Neuroscience}, vol.~15, no.~1, pp.~154--155, 2003.

\bibitem{kempter1999hebbian}
R.~Kempter, W.~Gerstner, and J.~L. Van~Hemmen, ``Hebbian learning and spiking
  neurons,'' {\em Physical Review E}, vol.~59, no.~4, p.~4498, 1999.

\bibitem{singer1995visual}
W.~Singer and C.~M. Gray, ``Visual feature integration and the temporal
  correlation hypothesis,'' {\em Annual review of neuroscience}, vol.~18,
  no.~1, pp.~555--586, 1995.

\bibitem{gupta2016oscillatory}
N.~Gupta, S.~S. Singh, and M.~Stopfer, ``Oscillatory integration windows in
  neurons,'' {\em Nature communications}, vol.~7, no.~1, pp.~1--10, 2016.

\bibitem{rutishauser2010human}
U.~Rutishauser, I.~B. Ross, A.~N. Mamelak, and E.~M. Schuman, ``Human memory
  strength is predicted by theta-frequency phase-locking of single neurons,''
  {\em Nature}, vol.~464, no.~7290, pp.~903--907, 2010.

\bibitem{hintzman1984minerva}
D.~L. Hintzman, ``Minerva 2: A simulation model of human memory,'' {\em
  Behavior Research Methods, Instruments, \& Computers}, vol.~16, no.~2,
  pp.~96--101, 1984.

\bibitem{kanerva1988sparse}
P.~Kanerva, {\em Sparse distributed memory}.
\newblock MIT press, 1988.

\bibitem{humphreys1989different}
M.~S. Humphreys, J.~D. Bain, and R.~Pike, ``Different ways to cue a coherent
  memory system: A theory for episodic, semantic, and procedural tasks.,'' {\em
  Psychological Review}, vol.~96, no.~2, p.~208, 1989.

\bibitem{pollack1990recursive}
J.~B. Pollack, ``Recursive distributed representations,'' {\em Artificial
  Intelligence}, vol.~46, no.~1-2, pp.~77--105, 1990.

\end{thebibliography}
\bibliographystyle{ieeetr}

\end{document}